\documentclass[printer]{aa}
\usepackage{graphicx}
\usepackage{txfonts}
\def\lesssim{\mathrel{\hbox{\rlap{\hbox{\lower4pt\hbox{$\sim$}}}\hbox{$<$}}}}
\def\gtrsim{\mathrel{\hbox{\rlap{\hbox{\lower4pt\hbox{$\sim$}}}\hbox{$>$}}}}
\newcommand{\mincir}{\raise -2.truept\hbox{\rlap{\hbox{$\sim$}}\raise5.truept
\hbox{$<$}\ }}
\newcommand{\magcir}{\raise -2.truept\hbox{\rlap{\hbox{$\sim$}}\raise5.truept
\hbox{$>$}\ }}
\newcommand{\siml}{\raise -2.truept\hbox{\rlap{\hbox{$\sim$}}\raise5.truept
\hbox{$<$}\ }}
\newcommand{\simg}{\raise -2.truept\hbox{\rlap{\hbox{$\sim$}}\raise5.truept
\hbox{$>$}\ }}
\newcommand{\be}{\begin{equation}}
\newcommand{\ee}{\end{equation}}
\newcommand{\ba}{\begin{eqnarray}}
\newcommand{\ea}{\end{eqnarray}}
\newcommand {\h} {Mpc$\;$}

\newcommand {\hhh} {\;\mathrm{Mpc}}
\newcommand {\hh} {Mpc}
\newcommand {\ks} {km~s$^{-1} \;$}
\newcommand {\kss} {km~s$^{-1}$}
\newcommand {\m} {$M_{\odot} \;$}

\newcommand {\mqui} {$\times 10^{15}M_{\odot} \;$}
\newcommand {\mquii} {$\times 10^{15}M_{\odot}$}

\newcommand{\degree}{\ensuremath{\mathrm{^\circ}}}
\newcommand{\arcm}{\ensuremath{\mathrm{^\prime}}}
\newcommand{\arcs}{\arcm\hskip -0.1em\arcm}
\newcommand{\dotarcs}{\,\rlap{\hbox{$\mathrm{^\prime\hskip-0.1em^\prime}$}}{\hbox{$.$}}\,}

\newcommand{\dotsec}{\,\rlap{\hbox{$\mathrm{^s}$}}{\hbox{$.$}}\,}
\begin{document}
   \title{Internal dynamics of the radio halo cluster
A2744}

\author{W. Boschin\inst{1,2} 
\and M. Girardi\inst{2,3}
\and M. Spolaor\inst{2}
\and R. Barrena\inst{4}}
   \offprints{W. Boschin; e.mail: boschin@tng.iac.es}

\institute{
Fundaci\'on Galileo Galilei - INAF, C/Alvarez de Abreu 70, E-38700 Santa Cruz de La Palma, Canary Islands, Spain\\
\and
Dipartimento di Astronomia, Universit\`{a} degli Studi di Trieste, via Tiepolo 11, I-34131 Trieste, Italy\\
\and
INAF -- Osservatorio Astronomico di Trieste, via Tiepolo 11, I-34131  Trieste, Italy\\
\and 
Instituto de Astrofisica de Canarias, C/Via Lactea s/n, E-38200 La Laguna, Tenerife, Canary Islands, Spain\\
}

\date{Received / accepted }

\abstract {} {We present a detailed dynamical analysis of the rich
galaxy cluster A2744, containing a powerful diffuse radio halo.}{Our
analysis is based on redshift data for 102 galaxies, part of them
recovered from unexplored spectra in the ESO archive.  We combine
galaxy velocity and position information to select the cluster members
and determine global dynamical properties of the cluster. We use a
variety of statistical tests to detect possible substructures.}  {We
find that A2744 appears as a well isolated peak in the redshift space
at $\left<z\right>=0.306$, which includes 85 galaxies recognized as
cluster members.  We compute the line--of--sight (LOS) velocity
dispersion of galaxies, $\sigma_{\rm V}=1767_{-99}^{+121}$ \kss, which
is significantly larger than what is expected in the case of a relaxed
cluster with an observed X--ray temperature of 8 keV. We find evidence
that this cluster is far from dynamical equilibrium, as shown by the
non--Gaussian nature of the velocity distribution, the presence of a
velocity gradient and a significant substructure.  Our analysis shows
the presence of two galaxy clumps of different mean LOS velocities
$\Delta V \sim 4000$ \kss.  We detect a main, low--velocity clump with
$\sigma_{\rm V}\sim$ 1200-1300 \ks and a secondary, high--velocity
clump with $\sigma_{\rm V}=$ 500-800 \kss and located in the S--SW
cluster region. We estimate a cluster mass within 1 \h of 1.4--2.4
\mquii, depending on the model adopted to describe the cluster
dynamics.}  {Our results suggest a merging scenario of two clumps with
a mass ratio of 3:1 and a LOS impact velocity of $\Delta V_{\rm rf}
\sim 3000$ \kss, likely observed just after the core passage. The
merging is occuring roughly in the NS direction with the axis close to
the LOS.  This scenario agrees with that proposed on the basis of
recent Chandra results in its general lines although suggesting a
somewhat more advanced merging phase. Our conclusions support the view
of the connection between extended radio emission and energetic
merging phenomena in galaxy clusters.\thanks{Table 1 is only available
in electronic form at the CDS via anonymous ftp to cdsarc.u-strasbg.fr
(130.79.128.5) or via http://cdsweb.u-strasbg.fr/cgi-bin/qcat?J/A+A/}}


\authorrunning{Boschin et al.}
\titlerunning{Internal dynamics of A2744} 
\maketitle

%______________________________________________________________________

\section{Introduction}

Clusters of galaxies are now recognized to be not simple relaxed , but
rather evolving via merging processes in a hierarchical fashion from
poor groups to rich clusters. Much progress has been made in recent
years in the observations of the signatures of the merging processes
(see Feretti et al. \cite{fer02a} for a general review). The presence
of substructures is indicative of a cluster in an early phase of the
process of dynamical relaxation or of secondary infall of clumps into
already virialized clusters.  The attempts to measure the substructure
occurrence rate for a large sample of galaxies give values of about
$50\%$ in both optical and X--ray data (e.g., Geller \& Beers
\cite{gel82}; Mohr et al. \cite{moh96}; Girardi et al. \cite{gir97};
Kriessler \& Beers \cite{kri97}; Jones \& Forman \cite{jon99};
Schuecker et al. \cite{sch01}).

A new aspect of these investigations is the possible connection of
cluster mergers with the presence of extended, diffuse radio sources,
halos and relics.  The synchrotron radio emission of halos and relics
demonstrates the existence of large--scale cluster magnetic fields, of
the order of 0.1--1 $\mu$G, and of widespread relativistic particles
of energy density 10$^{-14}$ -- 10$^{-13}$ erg cm$^{-3}$.  The
difficulty in explaining radio halos arises from the combination of
their large size, more than 1 $h_{50}^{-1}$ Mpc, and the short
synchrotron lifetime of relativistic electrons (e.g., Giovannini \&
Feretti \cite{gio02}).  Cluster mergers were suggested to provide the
large amount of energy necessary for electron reacceleration and
magnetic field amplification (Feretti \cite{fer99}; Feretti
\cite{fer02}; Sarazin \cite{sar02}). However, the question is still
debated since diffuse radio sources are quite uncommon and only
recently can we study these phenomena on the basis of sufficient
statistics ($\sim 30$ clusters up to $z\sim 0.3$, e.g., Giovannini et
al. \cite{gio99}; see also Giovannini \& Feretti \cite{gio02} and
Feretti \cite{fer05}).

Growing evidence of a connection between diffuse radio emission and
cluster merging is based on X--ray data (e.g., B\"ohringer \&
Schuecker \cite{boh02}; Buote \cite{buo02}). Studies based on a large
number of clusters have found a significant relation between the radio
and the X--ray surface brightness (Govoni et al. \cite{gov01a},
\cite{gov01b}) and connections between the presence of
radio halos/relics and an irregular and bimodal X--ray surface
brightness distribution (Schuecker et al. \cite{sch01}).  New
unprecedented insights into merging processes in radio clusters are
offered by Chandra and XMM--Newton observations (e.g., Markevitch \&
Vikhlinin \cite{mar01}; Markevitch et al. \cite{mar02}; Fujita et
al. \cite{fuj04}; Henry et al. \cite{hen04}; Kempner \& David
\cite{kem04}).

Optical data are a powerful way to investigate the presence and the
dynamics of cluster mergers (e.g., Girardi \& Biviano \cite{gir02}),
too.  The spatial and kinematical analysis of member galaxies allows
us to detect and measure the amount of substructure and to identify
and analyze possible pre--merging clumps or merger remnants.  This
optical information is complementary to X--ray information since
galaxies and ICM react on different time scales during a merger (see
numerical simulations by Roettiger et al. \cite{roe97}).
Unfortunately, to date optical data has been lacking or poorly
exploited.  The sparse literature concerns few individual clusters
(e.g., Colless \& Dunn \cite{col96}; G\'omez et al. \cite{gom00};
Barrena et al. \cite{bar02}; Mercurio et al. \cite{mer03}; Boschin et
al. \cite{bos04}).

Abell cluster 2744, also known as AC 118, is a rich, X--ray luminous,
hot cluster at moderate redshift: richness class =3 (Abell et
al. \cite{abe89}); $L_\mathrm{X}$(0.1--2.4 keV) = 22.05 $\times\,
10^{44} \ h_{50}^{-2}$ erg\ s$^{-1}$ (Ebeling et al. \cite{ebe96});
$T_\mathrm{X}=7.75_{-0.53}^{+0.59}$ (Allen \cite{all98}); z = 0.308
(Couch \& Newell \cite{cou84}).  Abell (\cite{abe58}) classified the
spatial distribution of its galaxies as ``regular'', but it has no
dominant bright galaxy or galaxies, so its Bautz--Morgan class is III
(Bautz \& Morgan \cite{bau70}).  Moreover, A2744 is known for an
excess of blue galaxies (Couch \& Newell \cite{cou84}), commonly known
as Butcher--Oemler effect (Butcher \& Oemler \cite{but78}).  Moreover,
A2744 hosts one of the most luminous known radio halos which covers
the central cluster region with a radius of $\sim 1$ $h_{50}^{-1}$
\hh, as well as a large radio--relic at a distance of about 2
$h_{50}^{-1}$ \h NE of the cluster center (Giovannini et
al. \cite{gio99}; Govoni et al. \cite{gov01a}, \cite{gov01b}).

The first indirect suggestion for a complex internal structure in
A2744 came from the high value of the LOS velocity dispersion of
member galaxies, $\sigma_{\rm V}\sim 1900$ \ks (Couch \& Sharples
\cite{cou87}). Another suggestion came from the discrepancy between
the gravitational shear amplitude and the velocity dispersion (Smail
et al. \cite{sma97}) and between the X--ray mass and strong lensing
mass (Allen \cite{all98}).  More direct evidences was found in
subsequent years.  In their study of internal dynamics of a large
cluster sample, Girardi \& Mezzetti (\cite{gir01}, hereafter GM01)
found two peaks in the velocity distribution of A2744, but these peaks
are so strongly superimposed that the authors questioned their
separation and classified A2744 as a cluster with uncertain dynamics.
ROSAT--PSPC data showed a subcluster 2.5\arcmin$\,$NW of the cluster
center (Govoni et al. \cite{gov01b}), where the galaxy distribution
seems to show an overdensity (Andreon \cite{and01}, based on deep
K--band photometry). This NW subcluster is also evident in
XMM--Newton data analyzed by Zhang et al. (\cite{zha04}), who measure
a negative temperature gradient moving from the center to the
outskirts of the cluster. Finally, the recent paper by Kempner \&
David (\cite{kem04}, hereafter KD04) used Chandra X--ray data to show
strong evidence for an ongoing major merger, detecting two secondary
X--ray peaks in the central cluster region. In particular, they
identified these peaks with the cool cores of two subclusters merging
in the NS direction, close to the LOS and suggested that the merging
is responsible for the radio halo.

In view of the above Chandra X--ray results and the availability of
unexplored spectroscopic data in the ESO archive we study the internal
dynamics of A2744 analyzing data of its member galaxies.

This paper is organized as follows.  We present the new redshift data
and the galaxy catalog in Sect.~2.  We analyze the internal dynamics
and detect substructures in Sects.~3 and 4, and the 2D galaxy
distribution in Sect.~5.  We discuss the dynamical status of A2744 in
Sect.~6 and summarize our results in Sect.~7.

Unless otherwise stated, we give errors at the 68\% confidence level
(hereafter c.l.).  Throughout the paper, we assume a flat cosmology
with $\Omega_{\rm m}=0.3$, $\Omega_{\Lambda}=0.7$, and $H_0=70$ \ks
Mpc$^{-1}$. For this cosmological model, 1 arcmin corresponds to $271$
kpc at the cluster redshift.

\section{Redshift data and galaxy catalog}

We considered multi--object spectroscopic data stored in ESO archive
acquired using NTT--EMMI in 1998 (2 MOS masks) and 1999 (another 2 MOS
masks).

Reduction of spectroscopic data was carried out with
IRAF\footnote{IRAF is distributed by the National Optical Astronomy
Observatories, which are operated by the Association of Universities
for Research in Astronomy, Inc., under cooperative agreement with the
National Science Foundation.}.  Radial velocities were determined
using the cross--correlation technique (Tonry \& Davis \cite{ton79})
implemented in the RVSAO package (developed at the Smithsonian
Astrophysical Observatory Telescope Data Center).  Each spectrum was
correlated against six templates for a variety of galaxy spectral
types: E, S0, Sa, Sb, Sc, Ir (Kennicutt \cite{ken92}).  The template
producing the highest value of $\cal R$, i.e., the parameter given by
the task XCSAO and related to the signal--to--noise of the correlation
peak, was chosen.  In two cases of galaxies with emission lines, we
took the EMSAO redshift as a more reliable estimate of the redshift.

We succeeded in obtaining 78 galaxy redshifts.  

\begin{figure*}[!htb]
\centering
\resizebox{\hsize}{!}{\includegraphics{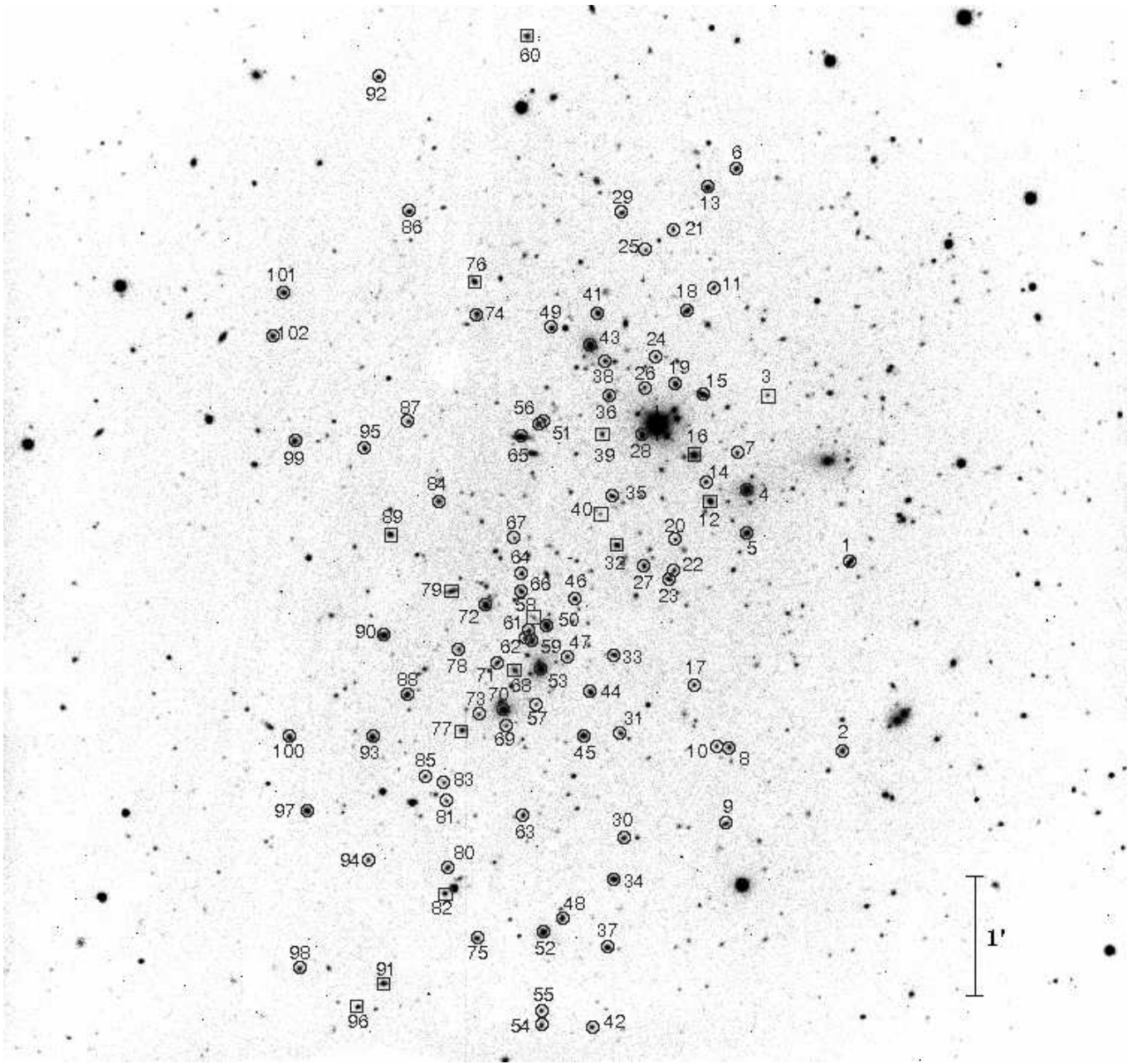}}
\caption{NTT R--band image of A2744 (north at the top and east to the
left) taken from the ESO archive (see also Busarello et
al. \cite{bus02}).  Galaxies with successful velocity measurements are
labeled as in Table~1. Circles and boxes indicate cluster member and
non member galaxies, respectively.}
\label{figimage}
\end{figure*}

\begin{figure*}[!htb]
\centering
\resizebox{\hsize}{!}{\includegraphics{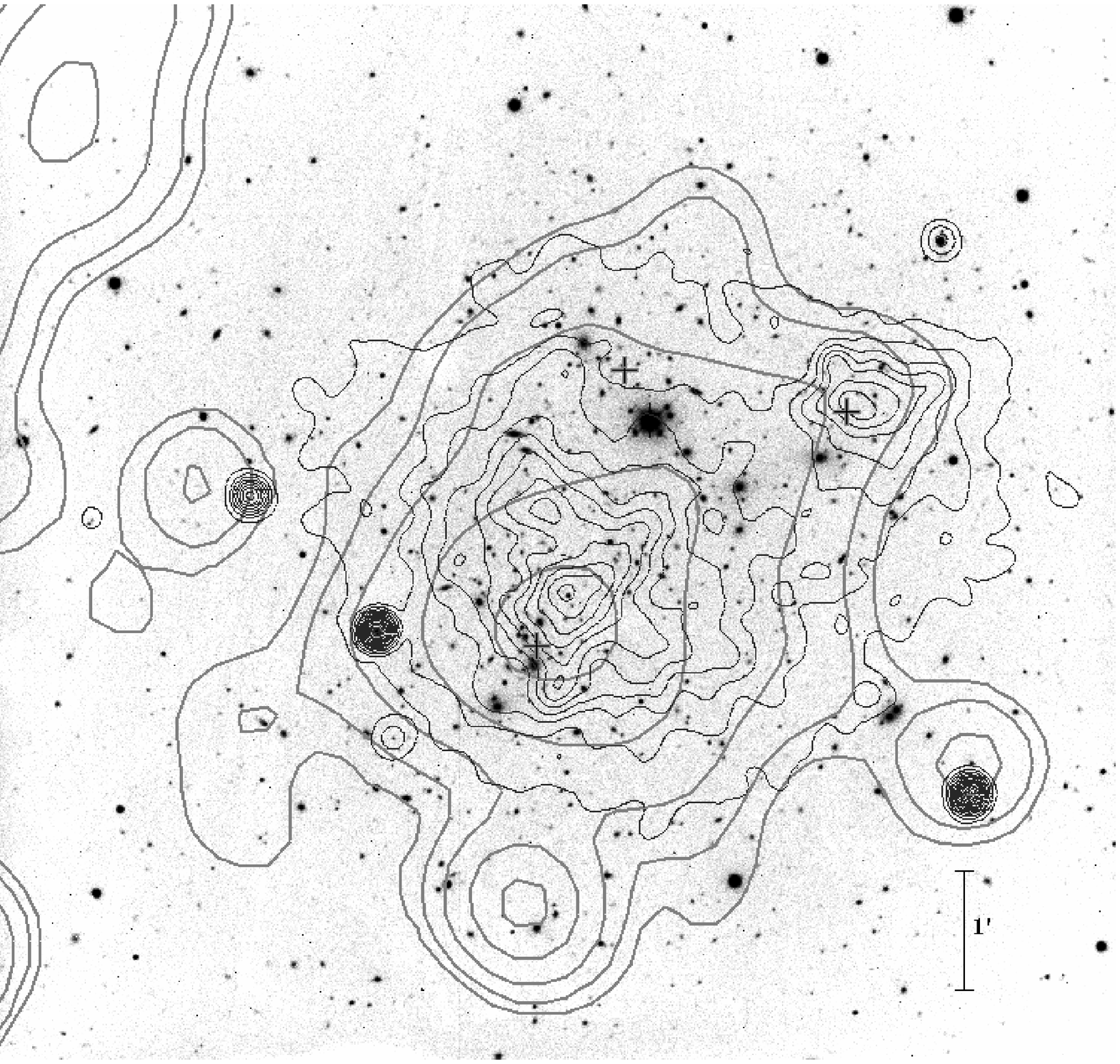}}
\caption{NTT R--band image of the cluster A2744 with, superimposed,
the contour levels of the Chandra archival image ID 2212 (thin black
contours; built considering X--ray photons in the energy range
0.5--2.0 keV) and of a VLA radio image at 1.4 GHz (thick gray
contours; courtesy of F. Govoni, see Govoni et al. \cite{gov01b}). The
X--ray image shows the central and the NW peaks, as well as the north
and south secondary peaks (see also KD04). The radio image shows the
powerful radio halo and (partially) the NE relic.  Crosses indicate
the position of the three peaks detected in the 2D galaxy distribution
of the likely cluster members according to the photometric redshifts
(see our Fig.\ref{figk2db}). North is at the top and east to the
left.}
\label{figisofote}
\end{figure*}

Multiple redshift determinations were available for a few galaxies: a
double determination for seven galaxies, a triple determination for
one galaxy, and a quadruple determination for one galaxy.  This
allowed us to obtain a more rigorous estimate for the redshift errors
since the nominal errors as given by the cross--correlation are known
to be smaller than the true errors (e.g., Malumuth et al.
\cite{mal92}; Bardelli et al. \cite{bar94}; Ellingson \& Yee
\cite{ell94}; Quintana et al. \cite{qui00}). Thus, we fitted the first
determination vs. the second one by using a straight line and
considering errors in both coordinates (e.g., Press et
al. \cite{pre92}). The fitted line agrees with the one to one
relation, but, when using the nominal cross--correlation errors, the
small value of $\chi^2$--probability indicates a poor fit, suggesting
that the errors are underestimated.  Only when nominal errors are
multiplied by a factor of $\sim 2.3$ can the observed scatter be
explained. Therefore, hereafter it is assumed that the true errors are
larger than the nominal cross--correlation errors by a factor of
2.3. The same value was obtained by Boschin et al. (\cite{bos04}) for
another cluster at moderate redshift.

Table~1 lists new radial velocity ${V_{\rm NTT}}=cz_{\sun}$
determinations we obtained for 66 galaxies (see also
Fig.~\ref{figimage}). The median error in radial velocity $\Delta
V_{\rm NTT}$ is $\sim 90$ \kss.  For the nine galaxies with multiple
redshift determinations, we considered the average of the
determinations and the corresponding error. Two galaxies (ID76 and 90)
show the presence of emission lines (OIIIb and H$\alpha$ in our
spectra). The galaxy ID90 also exhibits a powerful X--ray emission
(see Fig.~\ref{figisofote}).

We then considered the published redshifts, i.e. 40 galaxy redshifts
by Couch \& Sharples (\cite{cou87}, $\Delta V=100$ \kss) and 40 galaxy
redshifts by Couch et al. (\cite{cou98}, $\Delta V=90$ \kss). These
galaxies belong to the catalog of Couch \& Newell (\cite{cou84}) and
are numbered with the prefix ``CN''. Four galaxies are listed in both
the studies (CN219, CN258, CN310, CN360).  We cross--matched our
galaxies with those of the above authors: 14/30 galaxies of Couch \&
Sharples (\cite{cou87}) / Couch et al. (\cite{cou98}) were matched.
For common galaxies we repeated the straight line fitting--procedure
above separately for each study.  We found that the fitted lines agree
with the one to one relation and that adopted errors are able to
explain the observed scatters.  We added to our sample 26/10 galaxies
coming from Couch \& Sharples (\cite{cou87}) / Couch et
al. (\cite{cou98}).  For the galaxies with redshift determinations
from NTT data and these authors, we considered the average of the
determinations and the corresponding error.

Our final spectroscopic catalog consists of 102 galaxies sampling
$\sim 5\arcmin \times 8\arcmin$ in the central cluster region (see
Table~1 and Fig.~\ref{figimage}). It covers the region with
strong X--ray emission (with the exception of the NW peak) and the
region with the extended radio halo, but not the region with the
radio relic toward the NE (see Fig.~\ref{figisofote}).

Table~1 lists the velocity catalog (see also Fig.~\ref{figimage}):
identification number of each galaxy, ID (Col.~1); right ascension and
declination, $\alpha$ and $\delta$ (J2000, Col.~2); heliocentric
radial velocities obtained in this paper from NTT data, ${V_{\rm
NTT}}=cz_{\sun}$ (in \kss, Col.~3) with assumed errors, $\Delta
{V_{\rm NTT}}$, i.e., the nominal ones given by the cross--correlation
technique multiplied by 2.3 (Col.~4); heliocentric radial velocities
$V$ of the final catalog with assumed errors $\Delta V$ (Cols.~5 and
6); CN number for galaxies with previous redshift determination by
Couch \& Sharples (\cite{cou87}) and/or Couch et al. (\cite{cou98}) in
Column~7.\footnote{In reporting data of Couch et al. (\cite{cou98})
from their Table~3 we corrected a few typos (Couch, private
communication): the declination of CN97, CN130, CN340. Moreover galaxy
CN244 of Couch et al (\cite{cou98}) corresponds to CN258 of Couch \&
Sharples (\cite{cou87}).}

\section{Analysis of the whole system}

\subsection{Member selection}

The identification of cluster members proceeds in two steps, following
a procedure already used for nearby and medium--redshift clusters
(Fadda et al. \cite{fad96}; Girardi et al. \cite{gir96}; GM01).

\begin{figure}
\centering
\resizebox{\hsize}{!}{\includegraphics{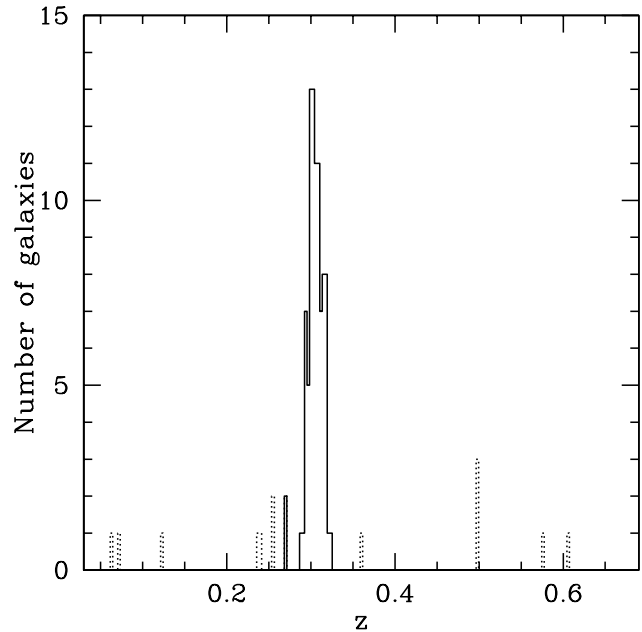}}
\caption
{Redshift galaxy distribution. The solid--line histogram 
refers to galaxies assigned to the cluster peak according to
the adaptive--kernel reconstruction method.}
\label{figden}
\end{figure}

First, we perform the cluster--member selection in velocity space by
using only redshift information. We apply the adaptive--kernel method
(Pisani \cite{pis93}, \cite{pis96}) to find the significant ($>99\%$
c.l.)  peaks in the velocity distribution.  This procedure detects
A2744 as a well isolated peak at $z=0.305$ assigning 89 galaxies
considered as candidate cluster members (see Fig.~\ref{figden}). Of
the non--member galaxies, seven and six are foreground and background
galaxies, respectively.

\begin{figure}
\centering 
\resizebox{\hsize}{!}{\includegraphics{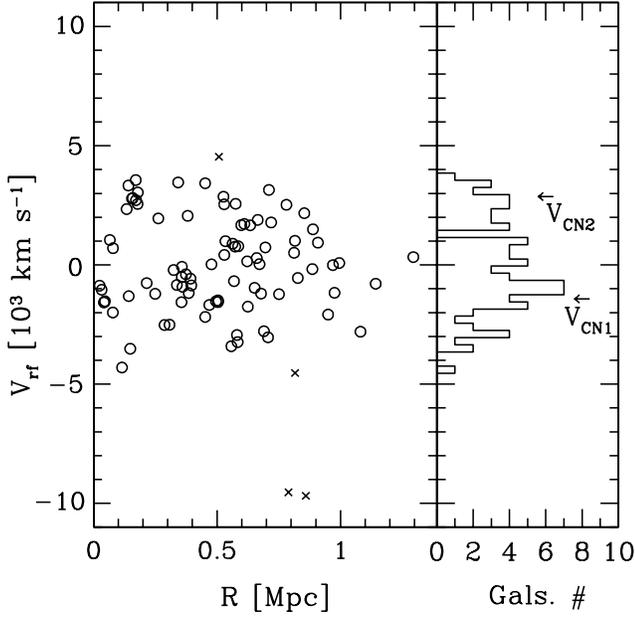}}
\caption
{ {\em Left panel}: rest--frame velocity vs. projected clustercentric
distance for the 89 galaxies in the main peak
(Fig.~\ref{figden}); the application of the ``shifting gapper''
method rejects four galaxies (crosses) selecting 85 cluster
members (circles). {\em Right panel}: velocity distribution of the 85
member galaxies. Velocities of the two brightest galaxies in the
central cluster region (ID53/CN1 and ID70/CN2) are indicated.}
\label{figvd}
\end{figure}

All the galaxies assigned to the A2744 peak are analyzed in the second
step, which uses both position and velocity information.  We apply the
procedure of the ``shifting gapper'' by Fadda et al. (\cite{fad96}).
This procedure rejects galaxies that are too far in velocity from the
main body of galaxies and within a fixed bin that shifts along the
distance from the cluster center.  The procedure is iterated until the
number of cluster members converges to a stable value.  Following
Fadda et al. (\cite{fad96}) we use a gap of $1000$ \ks -- in the
cluster rest--frame -- and a bin of 0.6 \hh, or large enough to
include 15 galaxies.  As for the cluster center we consider the
position of the most significant peak in the galaxy distribution as
obtained by the 2D adaptive--kernel analysis
[R.A.=$00^{\mathrm{h}}14^{\mathrm{m}}21\dotsec04$,
Dec.=$-30\degree\,23\arcmin\,52\dotarcs4$ (J2000.0), see Sect.~5].
This center is very close to the brightest cluster galaxy in the
central cluster region (ID53/CN1) which is also the center of the
cluster potential well according to the weak lensing analysis by Smail
et al. (\cite{sma97}).  The shifting--gapper procedure rejects four
galaxies as non--members (see Fig.~\ref{figvd}): two of them are
obvious interlopers since dist $\sim 5000$ \ks from the main body of
galaxies; the other two are emission line galaxies (ID12 and ID76).

Our following analysis is thus based on 85 cluster members.

\subsection{Global properties}

By applying the biweight estimator to cluster members (Beers et
al. \cite{bee90}), we compute a mean cluster redshift of
$\left<z\right>=0.3061\pm 0.0006$ ($\left<V\right>=91754\pm 192)$.  We
estimate the LOS velocity dispersion, $\sigma_{\rm V}$, by using the
biweight estimator and applying the cosmological correction and the
standard correction for velocity errors (Danese et al. \cite{dan80}).
We obtain $\sigma_{\rm V}=1767_{-99}^{+121}$ \kss, where errors are
estimated with a bootstrap technique.

\begin{figure}
\centering
\resizebox{\hsize}{!}{\includegraphics{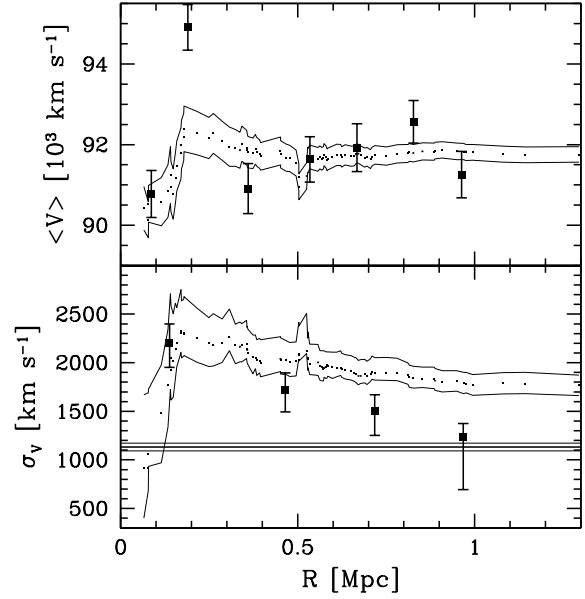}}
\caption
{Differential (big squares) and integral (small points) profiles of
mean velocity ({\em upper panel}) and LOS velocity dispersion ({\em
lower panel}).  As for the mean velocity/velocity dispersion, results
for seven/four annuli from the cluster center, each of 0.15/0.3 \hh,
are shown.  For the integral profiles, the mean and dispersion at a
given (projected) radius from the cluster center is estimated by
considering all galaxies within that radius. The error bootstrap bands
at the $68\%$ c.l. are shown.  In the lower panel, the horizontal line
represents the X--ray temperature with the respective 90 per cent
errors transformed in $\sigma_{\rm V}$ assuming the density--energy
equipartition between gas and galaxies, i.e.  $\beta_{\rm spec}=1$
(see text).}
\label{figprof}
\end{figure}

To evaluate the robustness of the $\sigma_{\rm V}$ estimate we analyze
the integral velocity dispersion profile (Fig.~\ref{figprof}).  The
value of the integral $\sigma_{\rm V}$ sharply varies in the internal
cluster region.  A similar behaviour is shown by the mean velocity
$\left<V \right>$ suggesting that a mix of galaxy clumps at different
redshifts is the likely cause of the high value of the velocity
dispersion rather than individual contaminating field--galaxies.  A
robust value of $\sigma_{\rm V}$ is reached in the external cluster
regions where the integral profile flattens, as found for most nearby
clusters (e.g., Fadda et al. \cite{fad96}).

If A2744 is in dynamical equilibrium (but see Sects.~3.3, 4, and 6.1)
one can compute global virial quantities.  Following the prescriptions
of GM01, we assume for the radius of the quasi--virialized region
$R_{\rm vir}=0.17\times \sigma_{\rm V}/H(z) = 3.7$ \h -- see their
eq.~1 after introducing the scaling with $H(z)$ (see also eq.~ 8 of
Carlberg et al. \cite{car97} for $R_{200}$). Thus the cluster is
sampled out to a radius of $R_{\rm out}=0.35\times R_{\rm vir}$.
Although the cluster is not sampled out to $R_{\rm vir}$, we can
compute the virial mass within $R_{\rm vir}$ following GM01.  We apply
the standard estimate (Limber \& Mathews \cite{lim60}) with the
surface term correction : $M=3\pi/2 \cdot \sigma_{\rm V}^2 R_{\rm
PV}/G-C$ (The \& White \cite{the86}; Girardi et al. \cite{gir98}),
where $\sigma_{\rm V}$ is the observed global velocity dispersion; the
radius $R_{\rm PV}$, equal to two times the (projected) harmonic
radius, is obtained assuming a King--like distribution, with
parameters typical of nearby/medium--redshift clusters ($R_{\rm
PV}=2.72\pm0.68$); the surface term correction $C$ is chosen to be
equal to the $20\%$ of the mass, a typical value for clusters (see
Carlberg et al. \cite{car97}; Girardi et al. \cite{gir98}).  We obtain
$M(<R_{\rm vir}=3.7 \hhh)=(7.4_{-2.0}^{+2.1})$\mquii.

\subsection{Velocity distribution}

We analyze the velocity distribution to look for possible deviations
from a Gaussian that could provide important signatures of complex
dynamics. For the following tests the null hypothesis is that the
velocity distribution is a single Gaussian.

We estimate three shape estimators, i.e. the kurtosis, the skewness,
and the scaled tail index (see, e.g., Beers et al. \cite{bee91}).  The
value of kurtosis (-0.85) shows evidence that the velocity
distribution differs from a Gaussian, being lighter--tailed, with a
c.l. of $\sim 95-99\%$ c.l. (see Table~2 of Bird \& Beers
\cite{bir93}).  The W--test (Shapiro \& Wilk \cite{sha65}) rejects the
null hypothesis of a Gaussian parent distribution at the $96.5\%$
c.l.

\begin{figure}
\centering 
\resizebox{\hsize}{!}{\includegraphics{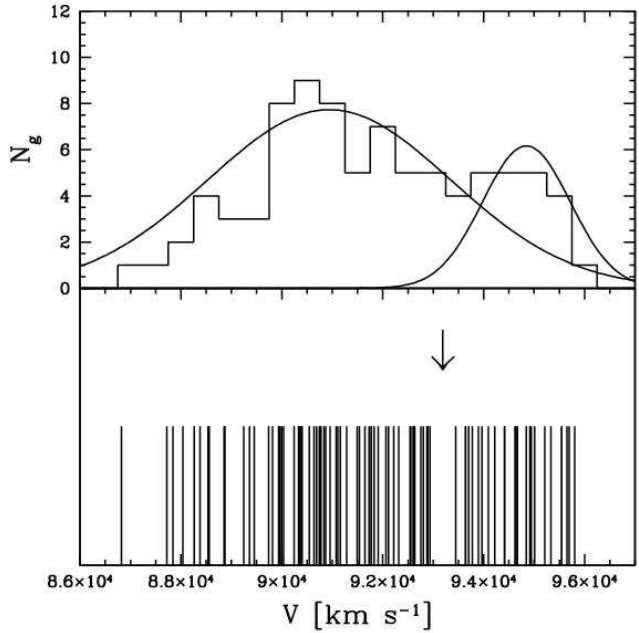}}
\caption
{Velocity distribution of radial velocities for the 85 cluster
members. {\em Lower panel}: stripe density plot where the arrow indicates
the position of the significant gap. The gap lies between 92936 and 93445
\kss.  {\em Upper panel}: velocity histogram with a binning of 500 \ks
with the Gaussians corresponding to KMMa and KMMb in
Table~\ref{tabsub}.}
\label{figstrip}
\end{figure}

We then investigate the presence of gaps in the distribution.  A
weighted gap in the space of the ordered velocities is defined as the
difference between two contiguous velocities, weighted by the location
of these velocities with respect to the middle of the data. We obtain
values for these gaps relative to their average size, precisely the
midmean of the weighted--gap distribution. We look for normalized gaps
larger than 2.25 since in random sampling of a Gaussian distribution they
arise at most in about $3\%$ of the cases, independent of the sample
size (Wainer and Schacht \cite{wai78}; see also Beers et
al. \cite{bee91}). One significant gap (3.074) in the ordered velocity
dataset is detected (see Fig.~\ref{figstrip}).  From low to high
velocities the dataset is divided in two subsets containing 61 and 24
galaxies.

\subsection{Velocity field}

The cluster velocity field may be influenced by the presence of
internal substructures, possible cluster rotation, and the presence of
other structures on larger scales, such as nearby clusters,
surrounding superclusters, and filaments. Each asymmetry effect could
produce a velocity gradient in the cluster velocity field.

\begin{figure}
\centering 
\resizebox{\hsize}{!}{\includegraphics{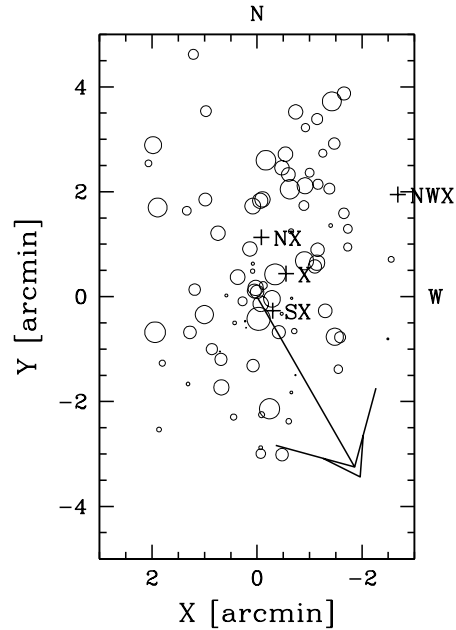}}
\caption
{Spatial distribution on the sky of the 85 member galaxies: the larger
the circle, the smaller the radial velocity.  The plot is
centered on the cluster center.  The arrow indicates the position
angle of the cluster gradient.  Crosses indicate the position of
X--ray peaks, i.e. the main cluster and the NW subcluster, and the
north and south secondary peaks (see Fig.~\ref{figisofote} and KD04).}
\label{figgrad}
\end{figure}

We estimate the direction of the velocity gradient performing a
multiple linear regression fit to the observed velocities with respect
to the galaxy positions in the plane of the sky (see also den Hartog
\& Katgert \cite{den96}; Girardi et al. \cite{gir96}). We find a
position angle on the celestial sphere of $PA=211_{-31}^{+19}$ degrees
(measured counter--clock--wise from north), i.e. higher--velocity
galaxies lie in the south--west region of the cluster, in agreement
with the visual impression of the galaxy distribution in
Fig.~\ref{figgrad}. To assess the significance of this velocity
gradient we perform 1000 Monte Carlo simulations by randomly shuffling
the galaxy velocities and for each simulation we determine the
coefficient of multiple determination ($RC^2$, see e.g., NAG Fortran
Workstation Handbook \cite{nag86}).  We define the significance of the
velocity gradient as the fraction of times in which the $RC^2$ of the
simulated data is smaller than the observed $RC^2$.  We find that the
velocity gradient is significant at the $98.2\%$ c.l.

\section{Detecting substructures}

The existence of correlations between positions and velocities of
cluster galaxies is a characteristic of real substructures.  We use
three different approaches to detect substructures in A2744 combining
velocity and position information.

\subsection{KMM analysis}

As a first approach, we attempt to separate in subsystems the whole
cluster starting from the velocity distribution.  The adaptive--kernel
method used in Sect.~3, which has the advantage of not requiring any a
priori shape for the subsystem research, is not able to distinguish
two peaks in the velocity distribution (see Fig.~\ref{fig040}, in the
next section). However, it finds a peak at $\sim 90850$ \ks and an
asymmetry at higher velocities, which suggests a hidden peak at
$94000-95000$ \kss.  To detect subsets in the velocity distribution we
then resort to the Kaye's mixture model (KMM) test (Ashman et
al. \cite{ash94}).

The KMM algorithm fits a user--specified number of Gaussian
distributions to a dataset and assesses the improvement of that fit
over a single Gaussian. In addition, it provides the
maximum--likelihood estimate of the unknown n--mode Gaussians and an
assignment of objects into groups. KMM is most appropriate in
situations where theoretical and/or empirical arguments indicate that
a Gaussian model is reasonable.  The Gaussian is valid in the case of
cluster velocity distributions, where gravitational interactions drive
the system toward a relaxed configuration with a Gaussian velocity
distribution.  However, one of the major uncertainties of this method
is the optimal choice of the number of groups for the partition.
Moreover, only in mixture models with equal covariance matrices for
all components (homoscedastic case; i.e., each group has the same
velocity dispersion) the algorithm converges, while this is not always
true for the heteroscedastic case (i.e., groups have different
velocity dispersion, see Ashman et al. \cite{ash94}, for further
details).

Our analysis of significant gaps suggests the presence of two
Gaussians. We present the results for the heteroscedastic KMM case
which we verify to be likely the corrected ipothesis.  We use the
results of the gap analysis to determine the first guess and we fit
two velocity groups around the guess mean velocities of $90\times10^3$
and $95\times10^3$ \kss. The algorithm fits a two--group partition at
the $97.5\%$ c.l., as obtained from the likelihood ratio test.

\begin{figure}
\centering 
\resizebox{\hsize}{!}{\includegraphics{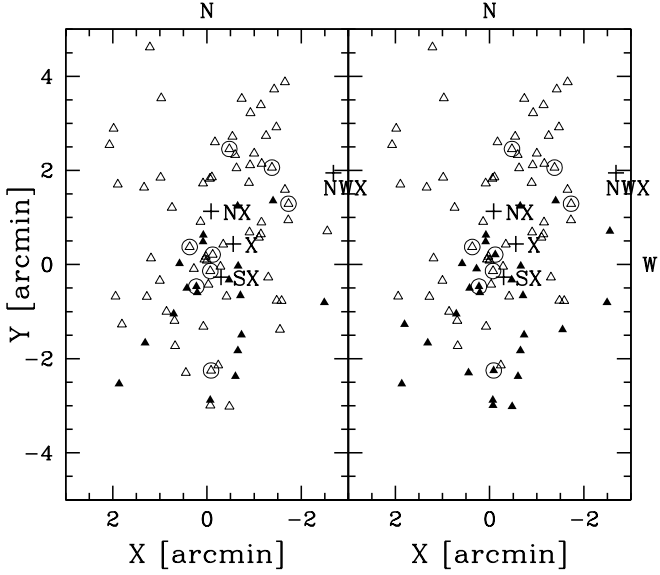}}
\caption
{Spatial distribution on the sky of the 85 member galaxies.  Open and
solid triangles indicate KMMa and KMMb groups ({\em left panel}) and
KMM3Da and KMM3Db groups ({\em right panel}).  Circles indicate the
nine red bright galaxies (see text). The plot is centered on the
cluster center.  Crosses indicate the position of X--ray peaks.}
\label{figkmm}
\end{figure}

Using the KMM algorithm, we assign member galaxies to individual
groups (66/19 for the low/high velocity group).  For both the groups,
we find that the Gaussian hypothesis of the velocity distribution is
acceptable, according to the W--test.  Moreover, the two groups appear
spatially segregated at the $98.7\%$ c.l. according to the 2D
Kolmogorov--Smirnov test -- hereafter 2DKS--test (Fasano \&
Franceschini \cite{fas87}, as implemented by Press et
al. \cite{pre92}).  Table~\ref{tabsub} shows the properties of the
corresponding groups KMMa and KMMb.  Fig.~\ref{figkmm} shows that the
galaxies of the high--velocity group populate mainly the SW cluster
region. Moreover, only one (ID70/CN2) of the eight red bright galaxies
belongs to the high--velocity group, where the red bright galaxies are
those with $R\le 19$ mag and $V-R>0.8$ (see also Sect.~5).

\addtocounter{table}{+1}
\begin{table}
        \caption[]{Results of kinematical and spatial analysis.}
         \label{tabsub}
                $$
         \begin{array}{l r l l}
            \hline
            \noalign{\smallskip}
            \hline
            \noalign{\smallskip}
\mathrm{Sample} & \mathrm{N_g} & \phantom{249}\mathrm{<V>}\phantom{249} & 
\phantom{24}\sigma_V^{\,\,\,\,\,\,\,\mathrm{a}}\phantom{24}\\
& &\phantom{249}\mathrm{km\ s^{-1}}\phantom{249} 
&\phantom{2}\mathrm{km\ s^{-1}}\phantom{24}\\
            \hline
            \noalign{\smallskip}
 
\mathrm{Whole\ system} & 85 &91754\pm192 &1767_{-99}^{+121}\\
\hline
\mathrm{KMM\ partitions}&&\\
\hline
\mathrm{KMMa} & 66 &90939\pm161 &1301_{-84}^{+130}\\
\mathrm{KMMb} & 19 &94846\pm110 &\phantom{1}462_{-55}^{+58}\\ 
\mathrm{KMM3Da}& 57 &90616\pm156 &1170_{-100}^{+116}\\
\mathrm{KMM3Db}& 28 &94370\pm163 &\phantom{1}841_{-108}^{+177}\\
\hline
\mathrm{2D\ subsamples}&&\\
\hline
\mathrm{R<0.4\ Mpc, lowV}& 23 &90537\pm180 &\phantom{1}841_{-169}^{+319}\\
\mathrm{R<0.4\ Mpc, highV}&12&94821\pm195&\phantom{1}631_{-103}^{+233}\\
\mathrm{R>0.4\ Mpc, N}   & 32 &91262\pm238 &1319_{-118}^{+176}\\
\mathrm{R>0.4\ Mpc, S}   & 18 &93081\pm476 &1940_{-317}^{+467}\\
\mathrm{R>0.4\ Mpc, W}   & 36 &91793\pm276 &1628_{-143}^{+206}\\
\mathrm{R>0.4\ Mpc, E}   & 14 &91676\pm500 &1779_{-189}^{+276}\\
\mathrm{R>0.4\ Mpc, NW}  & 24 &91311\pm278 &1325_{-146}^{+218}\\
\hline
\mathrm{Dressler\ \&\ Schectman\ substructure}&&\\
\hline
\mathrm{DS, lowV}& 17 &89932\pm202 &\phantom{1}798_{-104}^{+142}\\
\mathrm{DS, highV}&11&94046\pm269&\phantom{1}834_{-114}^{+157}\\
\hline
              \noalign{\smallskip}
            \hline
%            \noalign{\smallskip}
%            \hline
         \end{array}
$$
\begin{list}{}{}  
\item[$^{\mathrm{a}}$] We use the biweight and the gapper estimators by
Beers et al. (1990) for samples with $\mathrm{N_g}\ge$ 15 and with
$\mathrm{N_g}<15$ galaxies, respectively (see also Girardi et
al. \cite{gir93}).
\end{list}
         \end{table}

The Gaussian model for the 2D galaxy distribution is poorly supported
by theoretical and/or empirical arguments and, however, our galaxy
catalog is not complete down to a magnitude limit.  However, since the
3D diagnostics is in general the most sensitive indicator of the
presence of substructure (e.g., Pinkney et al. \cite{pin96}), we apply
the 3D version of the KMM software simultaneously using galaxy
velocity and positions. We use the galaxy assignment obtained in the
1D analysis to determine the first guess of the 3D analysis.  The
algorithm fits a two--group partition at the $97.7\%$ c.l. The
results for the two groups (KMM3Da, KMM3Db) are shown in
Table~\ref{tabsub} and Fig.~\ref{figkmm}.  The groups resulting from
the 3D analysis are similar to those resulting from the 1D analysis
for the difference in mean velocity and spatial distribution. The only
remark is that the 3D high--velocity group (KMM3Db) is a more
important structure with respect to the 1D high--velocity group
(KMMb), as indicated by the larger velocity dispersion and the
presence of three red bright galaxies.

\subsection{Analysis of different spatial regions}

In a second approach, we analyze the kinematical properties of galaxy
populations located in different spatial regions of the cluster.

\begin{figure}
\centering
\resizebox{\hsize}{!}{\includegraphics{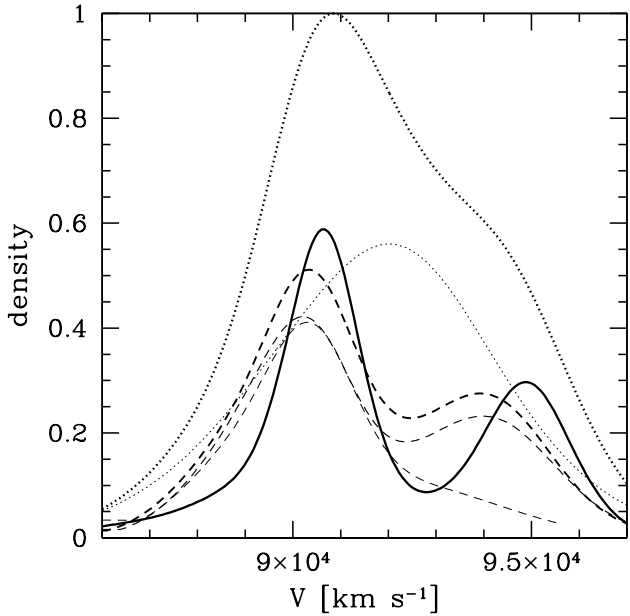}}
\caption
{The velocity galaxy density, as provided by the adaptive--kernel
reconstruction method for the whole cluster (thick--dotted line), the
region within 0.4 \h of the cluster center (thick--solid line) and
the region outside 0.4 \h (dotted line). Units on the y--axis are
normalized to the density of the highest peak in the whole
cluster. The plot also shows the velocity density for the galaxies
with Dressler--Schectman deviations (see Sect. 4.3) $\delta_{i}>1.6$,
1.7, and 1.8 (thick--dashed, dashed and thin--dashed lines,
respectively).}
\label{fig040}
\end{figure}

We consider the internal region within 0.4 \h of the cluster center
containing 35 galaxies. This central region allows the inclusion of
both secondary X--ray peaks and the exclusion of the NW X--ray
subcluster.  The velocity distribution of the central cluster region
shows the presence of two significant peaks, in agreement with the 1D
adaptive--kernel method results (see Fig.~\ref{fig040}): the main peak
at $V\sim 90600$ \ks and the secondary, with a relative density of
$\gtrsim 50\%$, at $V\sim 94900$ \kss. The overlap between the two
peaks in the galaxy assignment concerns only 2/35 galaxies and the
velocity separation is high ($\sim 4500$ \kss); thus, following the
criteria of Fadda et al. (\cite{fad96}), the two peaks are well
separable.

As for the external region, we divide it in a north and a south
region -- or alternatively in an east and a west region -- with
respect to the cluster center.  We also consider the NW sector for its
proximity to the NW X--ray subcluster.

Table~\ref{tabsub} lists the properties of these subsamples. Here we
point out interesting results.

The two peaks in the central region differ in mean velocity (at
$>99.99\%$ c.l. according to the means--test; Press et
al. \cite{pre92}) and both have a different mean velocity (at
$>99.9\%$ c.l.) and significantly smaller velocity dispersions than
the whole sample (at $99.9\%$ c.l. according to the F--test; Press et
al. \cite{pre92}). The galaxies of the two peaks are not spatially
segregated according to the 2DKS--test.  The south external region has
an high mean velocity, different from that of the north external
region at the $98.5\%$ c.l.

The above analysis confirms the presence of the two subsystems
found by the KMM method, with a secondary important high--velocity clump
in the south region, partially LOS aligned with the main clump in
the central cluster region.  This LOS alignment is the most likely
cause of the very high value of the velocity dispersion in the central
cluster region (see Fig.~\ref{figprof}).

\subsection{Dressler \& Schectman analysis}

Finally, we combine galaxy velocity and position information to
compute the $\Delta$--statistics devised by Dressler \& Schectman
(\cite{dre88}).  This test is sensitive to spatially compact
subsystems that have either an average velocity that differs from the
cluster mean, or a velocity dispersion that differs from the global
one, or both.  The subsystem which must be examined is not determined a
priori as in Sect.~4.2, but, for each $i$--th galaxy, the test considers the
subsample formed by this galaxy and the ten nearest neighbours and
computes a parameter, $\delta_{i}$, which gives the deviation of the
local kinematical parameters (velocity and velocity dispersion) from
the global cluster parameters. In the case of A2744 we find
$\Delta=130$ for the value of the parameter which gives the cumulative
deviation.  To compute the significance of the substructure we run 1000
Monte Carlo simulations, randomly shuffling the galaxy velocities, and
obtain a value of $98.9\%$.

\begin{figure}
\centering 
\resizebox{\hsize}{!}{\includegraphics{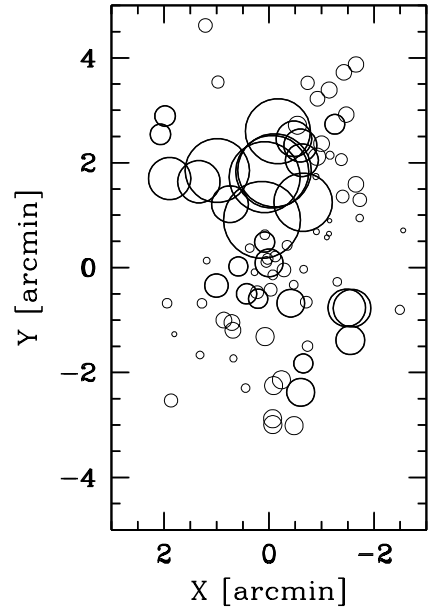}}
\caption
{Spatial distribution of cluster members, each marked by a circle: the
larger the circle, the larger is the deviation $\delta_{i}$ of the
local parameters from the global cluster parameters, i.e. there is
more evidence for substructure (according to the Dressler \& Schectman
test, see text).  The boldface circles indicate those with $\delta_{i}
\ge 1.7$ (see text).  The plot is centered on the cluster center.}
\label{figds}
\end{figure}

Fig.~\ref{figds} shows the distribution on the sky of all galaxies,
each marked by a circle: the larger the circle, the larger the
deviation $\delta_{i}$ of the local parameters from the global cluster
parameters, i.e. the higher the evidence for substructure.  This
figure provides information on the positions of substructures: some
galaxies with low velocity are the likely cause of large values of
$\delta_{i}$ in the N--NE region.

\begin{figure}
\centering 
\resizebox{\hsize}{!}{\includegraphics{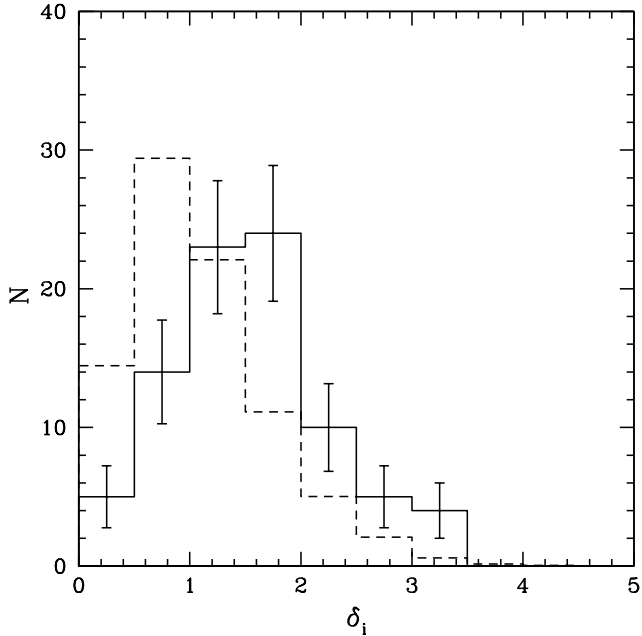}}
\caption
{The distribution of $\delta_{i}$ deviations of the
Dressler--Schectman analysis for the 85 member galaxies. The solid
line represents the observations, the dashed line the distribution for
the galaxies of simulated clusters, normalized to the observed
number.}
\label{figdeltai}
\end{figure}

To obtain further information, we resort to the technique developed by
Biviano et al. (\cite{biv02}), who used the individual
$\delta_{i}$--values of the Dressler \& Schectman method. The critical
point is to determine the value of $\delta_{i}$ that optimally
indicates galaxies belonging to substructure. To this aim we consider
the $\delta_{i}$--values of all 1000 Monte Carlo simulations already
used to determine the significance of the substructure (see above).
The resulting distribution of $\delta_{i}$ is compared to the observed
one. We find a difference at $P=99.98\%$ c.l. according to the
KS--test.  The ``simulated'' distribution is normalized to produce the
observed number of galaxies and compared to the observed distribution
in Fig.~\ref{figdeltai}: the latter shows a tail at large values. This
tail is populated by galaxies that presumably are in substructures.
For the selection of galaxies within substructures we choose the
threshold value of $\delta_{\rm th}=1.7$, since after the rejection of
the values $\delta_{i}>\delta_{\rm th}$, the observed and simulated
$\delta_{i}$--distributions are still different at $P\sim 99\%$.  With
this choice, 28 galaxies of the cluster are assigned to substructure.
They are located in the NE region, in the center and in the SW region
(see Fig.~\ref{figds}).  Moreover, their velocity distribution shows
two significant peaks at $\sim 90000$ \ks and $\sim 94000$ \ks (see
Fig.~\ref{fig040}). The results of the kinematical analysis of the
two galaxy--groups assigned to the two velocity peaks are listed in
Table~\ref{tabsub}.  The values of the velocities of the two peaks
suggest us that the Dressler--Schectman test detects (part of) the same
two galaxy-structures already detected in the above sections.

The two peaks in the velocity distribution are also found with a less
conservative threshold value for substructure galaxies (e.g., 36
galaxies with $\delta_{i}> \delta_{\rm th}=1.6$, $P\sim 95\%$), while
with a more conservative threshold only the $\sim 90000$ \ks peak is
detected (see Fig.~\ref{fig040}).

\section{2D galaxy distribution}

\begin{figure}
\centering
\includegraphics[width=7cm]{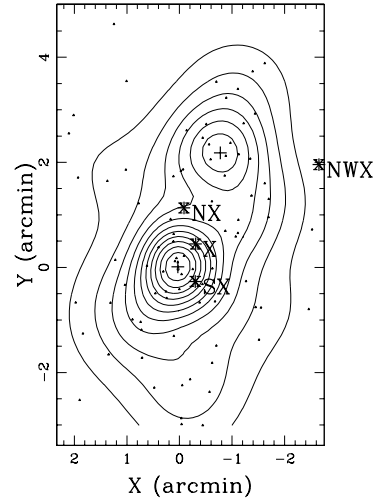}
\caption
{Spatial distribution on the sky and relative isodensity contour map
of 85 cluster members (spectroscopically confirmed), obtained with the
adaptive--kernel method.  The plot is centered on the cluster center.
Crosses and stars indicate the position of the galaxy and X--ray
peaks, respectively.}
\label{figk2da}
\end{figure}

By applying the 2D adaptive--kernel method to the position of A2744
galaxy members we find two very significant peaks ($>99.99\%$): the
densest peak (hereafter CG peak) with 43 galaxies lies on the cluster
center and the secondary (hereafter NG peak) with 33 galaxies lies to
the north and slightly NW
[R.A.=$00^{\mathrm{h}}14^{\mathrm{m}}17\dotsec3$
Dec.=$-30\degree\,21\arcmin\,42\arcs$ (J2000.0), see
Fig.~\ref{figk2da}].

Unfortunately our galaxy catalog has two limits that could bias the
above result: the small extension covered by spectroscopical data and
the magnitude incompleteness. To overcome these limits we consider the
catalog presented by Busarello et al. (\cite{bus02}). These authors
present photometric V--, R--, and I--band data in a catalog of 1206
sources, where the galaxy sample is complete to R=22.3 mag, and give
photometric redshifts for 459 sources for which additional U-- and
K--band photometry is available.

\begin{figure}
\centering
\includegraphics[width=7cm]{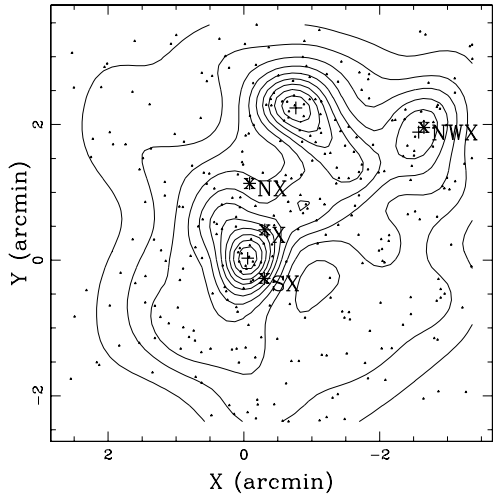}
\caption
{Spatial distribution on the sky and relative isodensity contour map
of 320 likely cluster members (according to the photometric redshifts
by Busarello et al. \cite{bus02}), obtained with the adaptive--kernel
method.  The plots are centered on the cluster center.  
Crosses and stars indicate the
position of the galaxy and X--ray peaks, respectively.}
\label{figk2db}
\end{figure}

Following the cluster member definition of Busarello et
al. (\cite{bus02}), we select all galaxies, i.e. sources with stellar
index $\le 0.9$, having photometric redshift such that the associated
$68\%$ percentile interval $\Delta z_{68}$ overlaps the range $z \in
[0.24,0.38]$. We obtain a catalog of 320 objects. Fig.~\ref{figk2db}
shows the result of the 2D adaptive--kernel method. We confirm the
presence of the peaks found with the spectroscopic members: 113
galaxies are assigned to the CG peak and 81 to the NG peak.  Moreover,
there is the presence of another significant peak (hereafter NWG peak)
with 37 galaxies at the position of the NW X--ray peak
[R.A.=$00^{\mathrm{h}}14^{\mathrm{m}}08\dotsec96$
Dec.=$-30\degree\,21\arcmin\,59\dotarcs6$ (J2000.0)].

\begin{figure}
\centering
\resizebox{\hsize}{!}{\includegraphics{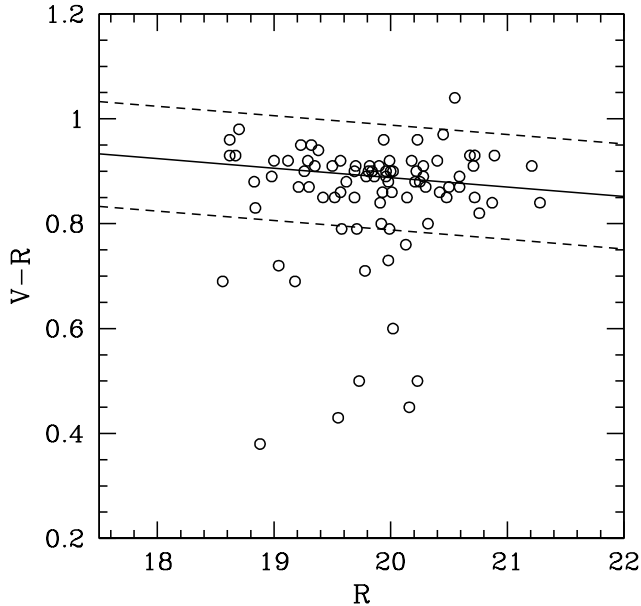}}
\caption
{V--R vs. R diagram for spectroscopically determined cluster galaxies.
The solid line gives the best--fit color--magnitude relation; the
dashed lines are drawn at $\pm$0.1 mag from the relation.  }
\label{figcm}
\end{figure}

\begin{figure}
\centering
\includegraphics[width=7cm]{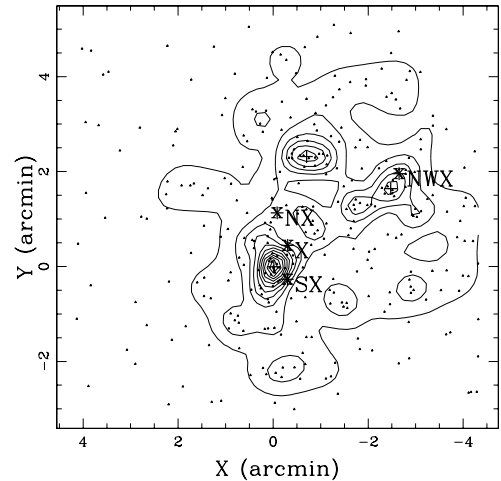}
\caption
{Spatial distribution on the sky and relative isodensity contour map
of 316 likely cluster members (according to the color--magnitude
relation) with $R\le 22.3$, obtained with the adaptive--kernel method.
The plots are centered on the cluster center. 
The stars
indicate the position of the X--ray peaks.
Crosses and stars indicate the
position of the galaxy and X--ray peaks, respectively.}
\label{figk2dc}
\end{figure}

Finally, we also consider the whole photometric galaxy catalog by
Busarello et al. (\cite{bus02}) after a selection through the
color--magnitude relation, which indicates early--type galaxy locus.
Specifically, out of the R--magnitude complete catalog we select 316
likely cluster members considering galaxies within 0.1 mag from the
$\rm{V}-\rm{R}=1.248-0.018\cdot \rm{R}$ relation, obtained by using a
2 sigma--clipping fitting procedure to our spectroscopic catalog of
member galaxies (see Fig.~\ref{figcm}). Several peaks are shown in
Fig.~\ref{figk2dc}, but there are only three peaks with significance
greater than the $99.99\%$ c.l.; they are those already detected
above: the central, the N, and the NW peaks (56, 40, and 22 galaxies,
respectively).

\section{Discussion}

We analyze the internal dynamics of A2744 on the basis of
spectroscopic data for 102 galaxies in a cluster region of $\sim
5\arcmin\times 8\arcmin$ (i.e., within a radius of $\sim 1.3$ \h from
the cluster center).  In particular, we present new redshifts for 66
galaxies, as recovered from spectra stored in the ESO archive. These
redshifts have allowed us to expand our knowledge with respect to
previous published redshifts, mainly in the northern cluster
region.

The value we obtain for the LOS velocity dispersion ($\sigma_{\rm
V}\sim 1800$ \kss, see also Couch \& Sharples \cite{cou87} and GM01)
is very high for clusters (see Fadda et al. \cite{fad96}; Mazure et
al. \cite{maz96}; GM01).  This global value of $\sigma_{\rm V}$ is
inconsistent with the average value of $T_{\rm X} \sim 8$ keV coming
from the X--ray analysis when assuming the equipartition of mass
density between gas and galaxies (i.e. $\beta_{\rm spec}
=2.4^{+0.5}_{-0.4}$ to be compared with $\beta_{\rm
spec}=1$\footnote{$\beta_{\rm spec}=\sigma_{\rm V}^2/(kT/\mu m_{\rm
p})$ with $\mu=0.58$ the mean molecular weight and $m_{\rm p}$ the
proton mass.}, see also Fig.~\ref{figprof}).

This suggests that the cluster is
far from dynamical equilibrium.  Indeed, we do find significant
signatures of the young dynamical status of A2744: the non--Gaussian nature
of the velocity distribution, the presence of a velocity gradient and
the detection of significant substructure according to the
Dressler--Schectman test.

\subsection{Clump detection}

In their study of the internal dynamics of a large cluster sample,
GM01 found two peaks in the velocity distribution of A2744 separated
by $\sim 4000$ \ks and having $\sigma_{\rm V}\sim 1100$ and 700 \kss,
respectively.  However, these peaks were so strongly superimposed that
the authors questioned their separation and classified A2744 as a
cluster with uncertain dynamics.  The present analysis, performed ad
hoc for A2744, supercedes the GM01 results.  The presence of two well
separated peaks ($\Delta V\sim 4000$ \kss) characterizes the central
cluster region, while the external cluster region is well described by
a one--peak velocity distribution. The presence of two galaxy
populations having very different mean velocities in the central
cluster region explains the high value of the LOS velocity dispersion
found there and the sharp change in the mean velocity (see
Fig.~\ref{figprof}).  Our result also explains the strong mass
discrepancy between X--ray and lensing mass estimates in central
cluster regions ($M_{\rm arc}/M_{\rm X}>3$, Allen \cite{all98}).  In
fact, substructure and LOS alignments of material toward the cluster
core likely enhances the mass determined from the lensing data (e.g.,
Bartelmann \& Steinmetz \cite{bar96}).

As for the identification of A2744 subsystems, using the KMM technique
we find that a bimodal distribution is a significantly better
description in the velocity space, as well as in the
position--velocity space, and we assign the galaxies to two
groups. The velocity difference is $\Delta V \sim 4000$ \ks in
agreement with that of the two peaks detected in the central region.
Accordingly, the galaxies belonging to substructure, as detected
through the Dressler--Schectman test, split into two populations of
different velocity.  All our results indicate the presence of two
galaxy--clumps, one at $\left<V\right>=90000-91000$ \ks and the other
at $\left<V\right>=94000-95000$ \kss.

The KMM analysis indicates that the low--velocity clump is the main
one (with $\sigma_{\rm V} \sim$ 1200 -- 1300 \kss) and that the
high--velocity clump is the secondary one (with $\sigma_{\rm
V}=500-800$ \kss). The value of $\sigma_{\rm V} \sim$ 1200 -- 1300 \ks
of the main clump corresponds to $T_{\rm X}\sim 9$ keV, in
agreement with the average X--ray temperature (Allen
\cite{all98}). According to the KMM results, the relative importance
of the secondary clump is largely uncertain.  Since our analysis of
the central cluster region within 0.4 \h supports an high relative
importance, hereafter we consider only the 3D KMM results.  Following
the same procedure of Sect.~3.2, the values of $\sigma_{\rm V}$ for
the two clumps allow us to compute virial masses separately, assuming
that each of the two clumps are virialized systems.  We obtain $M_{\rm
a}(<R_{\rm vir,a}=2.4 \hhh)=(2.2_{-0.6}^{+0.7})$\mqui and $M_{\rm
b}(<R_{\rm vir,b}=1.7 \hhh)=(0.8_{-0.3}^{+0.4})$\mqui for the 3D KMM
case.

\subsection{Cluster mass estimate}

In Sect.~3.2 we give the mass estimate assuming the whole cluster as
virialized, $M(<R_{\rm vir}=3.7 \hhh)=(7.4_{-2.0}^{+2.1})$\mquii.
Since the system is not virialized, but likely formed by bound
structures, this estimate might overestimate the true cluster mass
even by a factor two.  The crude addition of the mass of the two
clumps gives a value that is less than half, i.e. $\sim 3$
\mquii. However, this should be considered an underestimate of the
true cluster mass since it does not consider the NW external
subcluster as well as other infalling minor groups.  To obtain a more
meaningful mass estimate we use both the one--clump and two--clump
approaches to compute the mass values within a radius of 1 \hh, which
is well sampled by the spectroscopic data.  In the case of the
two--clump approach we consider the global cluster geometry as formed
by the two clumps at the cluster redshift assuming the center of A2744
for both the clumps (see the following section). To rescale the mass
within 1 \hh, we assume that each system is described by a King--like
mass distribution (see Sect.~3.2) or, alternatively, by a NFW profile
where the mass--dependent concentration parameter is taken from
Navarro et al. (\cite{nav97}) and rescaled by the factor 1+z (Bullock
et al. \cite{bul01}; Dolag et al. \cite{dol04}). We obtain $M(<R=1
\hhh)=1.8 - 2.4 \times 10^{15}$\m in the one--clump assumption and
$M(<R=1 \hhh)=1.4 - 1.5 \times 10^{15}$\m in the two--clump
assumption. Such values are comparable to those of very massive
clusters at lower redshifts (e.g., Girardi et al. \cite{gir98}; GM01).

\subsection{Merging scenario}

As discussed in the above section, we detect two clumps with the mass
ratio of about 3:1 and with a velocity separation of $\Delta V_{\rm
rf} \sim 3000$ \ks in the cluster rest--frame.  Cosmological
simulations suggest that high impact velocities might be occurring in
cluster mergers (e.g., Crone \& Geller \cite{cro95}).  The value of a
relative velocity of $\sim 3000$ \ks is predicted by N--body numerical
simulations of Pinkney et al. (\cite{pin96}) and is likely associated
with the phase of the core passage in a cluster--cluster merging (see
their Fig.~1 for a merging with a 3:1 mass ratio). In particular, the
extremely large LOS component of the velocity indicates that the
merging axis is close to the LOS.

From the observation of the cool wakes in the Chandra image, KD04
deduced that the merger is not occurring entirely along the LOS, but
must have at least a small transverse component. However, analyzing
the galaxy distribution they found no conclusive evidence for a
spatial segregation (i.e. only at the $\gtrsim 90\%$ c.l.).  Likely
due to our larger spectroscopic sample, we find definitive evidence
for spatial segregation as shown by the velocity gradient, the
segregation of KMM groups, and the analysis of the south external
region ($98-99\%$ c.l.). All our results agree in displacing the
high--velocity group to the S--SW region.

Although the two clumps are spatially segregated, their 2D structure
(i.e. the position of the respective centers and cores) is not
obvious. KD04 pointed out this difficulty as due to the small sampled
cluster region.  In spite of the advantage of using a somewhat larger
spectroscopic sample and the much more extended 2D sample of likely
members, we do not reach a definitive conclusion.

Apart from the NW peak (see below), our 2D analysis of the galaxy
distribution (Sect.~5) detects two significant peaks: the main peak in
the central cluster region (CG peak) and the secondary peak $\sim
2\arcmin$ toward N--NW (NG peak). The brightest red cluster galaxies,
which generally lie in cluster cores, are located in the same
overdense regions.  The segregation between the gas and galaxy
subclustering is extremely severe. The main X--ray peak lies between
CG and NG peaks, much closer to the CG peak and there is no spatial
correspondence between X--ray secondary peaks and 2D galaxy peaks.

As for the CG peak, the presence of two galaxy populations having
different mean velocities suggests that the cores of the two clumps
might be aligned along the LOS. However, the strong spatial
segregation of the two clumps suggests that they have not preserved
their original structure, but that one or likely both clumps are
strongly affected by the merger. In fact, the CG peak is not at the
center of the 2D distribution of the high--velocity clump, but rather
at the border. A few galaxies, in particular the most luminous/massive
and thus the most affected by dynamical friction, might be slowed down
by the interaction between the two clumps, while the other galaxies
pass by.  The presence in the CG peak of two very luminous galaxies
differing in velocity by 4500 \ks (ID53/CN1 and ID70/CN2) supports our
hypothesis: they might be the tracers of the two cores destined to
oscillate around the mean velocity for a long time (i.e., several
10$^9$ yr as suggested by numerical simulations; see, e.g., Nakamura
et al. \cite{nak95}). In fact, an high velocity difference between
two dominant galaxies is often suggestive of an energetic cluster
merger (e.g., Burns et al. \cite{bur95}) and indeed this process is
thought to be the cause of the formation of dumbbell galaxies in a few
merging clusters (e.g., Beers et al. \cite{bee92}; Flores et
al. \cite{flo00}).  Galaxies that are not slowed down by cluster
merging (collisionless galaxies) are likely those causing the velocity
gradient, i.e.  galaxies of the high--velocity clump located in the
S--SW region and galaxies of the low--velocity clump located in the
N--NE region. Since the intracluster medium is not relaxed, as shown
by the presence of the two secondary X--ray peaks, we suggest an early
merging phase.  If we are observing the merging just after the first
core passage, the low--velocity clump is moving north and the
high--velocity clump is moving south.

In the above scenario the identification of the north secondary X--ray
peak (XN) with the gas core of the low--velocity clump and of the
south secondary X--ray peak (XS) with the gas core of the
high--velocity clump -- see KD04 -- is still meaningful, in spite of
the segregation between the gas and galaxy subclustering.  In fact,
observations agree with the fact that (collisionless) galaxies, as
detected at the north and south cluster regions, precede the
(collisional) intracluster medium.

In summary, we suggest that the two galaxy clumps are in an advanced
phase of merging, just after the core passage, but still far from
having formed a dynamically relaxed cluster.  We agree with the
merging scenario proposed by KD04 in its general lines although
propose a somewhat more advanced merging phase (see their
Sect.~5.1). In fact, they assume that the gas has not yet decoupled
from the dark matter and galaxies, while we find a strong segregation
between the gas and galaxy subclustering.  Moreover, we agree with
KD04 about the importance of merging and the high velocity of the
impact, thus supporting their hypothesis that this merger in the
central cluster region is responsible for the powerful radio halo.

Indeed, the cluster structure described by our analysis is also more
complex than that suggested by Chandra data.  The NG peak shown by our
2D analysis, which has no counterpart in the X--ray distribution,
represents an additional feature in the above scenario.  The galaxies
located around this peak have kinematical properties similar to those
of the low--velocity clump.  However, we discard the hypothesis that
the NG peak represents the undisturbed core of the low--velocity main
clump due to its large distance from the main X--ray peak, likely
tracing the potential well of the just forming cluster. About the
nature of the NG peak, we propose two alternatives, both possible in
the above main merging scenario.  It might be a relic of a previous
merging involving the main clump (some galaxy structures might be
detectable in the host cluster for a long time, see Gonz\'alez--Casado
et al. \cite{gon94}) or a part of the core of the main clump that
survived the ongoing merging.  The second hypothesis is supported by
its location along the direction of the ridge A shown by the smoothed
X--ray surface brightness image of KD04, i.e. the ridge hosting the
north secondary X--ray peak (see their Fig.~2 and
Fig.~\ref{figisofote} in this paper).

As for the external cluster region, our 2D analysis finds a NW peak
(NWG peak) in the galaxy distribution corresponding to the NW X--ray
peak.  The galaxy peak is somewhat farther east than the X--ray peak,
in agreement with the NW clump moving from west to east (KD04) and
that collisionless faint galaxies precede the collisional intracluster
medium during the interaction.  The presence of an ongoing merging in
the $\sim$NS direction and a future merging in the $\sim$WE direction
suggests that A2744 lies in the intersection of filaments (e.g.,
cluster A521 analyzed by Arnaud et al. \cite{arn00}), which is the
natural place of rich cluster formation (e.g., Katz \& White
\cite{kat93}).  Spectral observations of galaxies in a large field
around A2744 would provide new insights into this issue.

\section{Summary \& conclusions}

We present the results of the dynamical analysis of the rich, X--ray
luminous and hot cluster of galaxies A2744, containing a powerful
diffuse radio halo.  The X--ray emission is known to have two
secondary cores in the central region and a peak in the NW external
region (Govoni et al. \cite{gov01b}; KD04). GM01 found two peaks in
the velocity distribution, but these peaks were so strongly
superimposed that the authors questioned their separation and
classified A2744 as a cluster with uncertain dynamics.  The present
analysis, performed ad hoc for A2744, supercedes the GM01 results.

Our analysis is based on velocities and positions of 102 galaxies in a
cluster region of $\sim 5\arcmin\times 8\arcmin$ (i.e., within a
radius of $\sim 1.3$ \h from the cluster center).  In particular, we
present new redshifts for 66 galaxies, as recovered from spectra
stored in the ESO archive.

We find that A2744 appears as a well isolated peak in the redshift
space at $\left<z\right>=0.306$, which includes 85 galaxies recognized
as cluster members. We compute the LOS velocity dispersion
of galaxies, $\sigma_{\rm V}=1767_{-99}^{+121}$ \kss, which is
significantly larger than what is expected in the case of a relaxed
cluster with an observed X--ray temperature of 8 keV.

We find evidence that this cluster is far from dynamical equilibrium,
as shown by:

\begin{itemize}

\item the non--Gaussian nature of the velocity distribution according to
different tests;

\item 
the presence of a velocity gradient at the $>98\%$ c.l.;

\item the presence of significant substructures at the $\sim99\%$ c.l.

\end{itemize}

To detect and analyze possible subsystems we use different methods
coupling galaxy position and velocity information: the KMM method, the
kinematical analysis of galaxy population located in different spatial
regions, and Dressler \& Schectman statistics.

Our analysis shows the presence of two galaxy clumps with:

\begin{itemize}

\item a difference in the mean velocities of $\Delta V \sim 4000$ \ks;

\item 
a value of $\sigma_{\rm V} \sim$ 1200 -- 1300 \ks for the main,
low--velocity clump and a value of $\sigma_{\rm V} = 500-800$ \ks for
the secondary, high--velocity clump;

\item 
a remarked spatial segregation, with the galaxies of the 
high velocity clump mainly populating the S--SW cluster region.

\end{itemize}

Our results suggest a merging scenario of two clumps with a mass ratio
of 3:1 and an high LOS impact velocity of $\Delta V_{\rm rf} \sim
3000$ \kss, likely observed just after the core passage. The merging
axis, close to the LOS, is roughly along the NS direction.  This
scenario supports recent Chandra results (KD04) although suggests a
somewhat more advanced merging phase.  For the final product of the
merger we estimate a mass within 1 \h of 1.4--2.4 \mquii, depending on
the model adopted to describe the cluster dynamics.  Such values are
comparable to those of very massive clusters at lower redshifts.

Our conclusion supports the view of the connection between extended
radio emission and energetic merging phenomena in galaxy clusters.

\begin{acknowledgements}
We thank Federica Govoni for the VLA radio contour levels she kindly
provided us. We thank the referee, Florence Durret, for her useful 
suggestions.\\
This publication is based on observations made with the NTT telescope
at the La Silla observatory (proposal IDs: 62.O--0369(A),
63.O--0257(B) and 64.O--0236(B)). Data are stored in the public ESO
archive. This publication also makes use of data obtained from the
Chandra data archive at the NASA Chandra X--ray center
(http://asc.harvard.edu/cda/).
\end{acknowledgements}

\end{document}